\documentclass[twocolumn, graphics, a4paper, floatfix,  superscriptaddress]{revtex4-2}

\usepackage[english]{babel}
\usepackage[letterpaper,top=2cm,bottom=2cm,left=1.2cm,right=1.2cm,marginparwidth=1.75cm]{geometry}

\usepackage{amsmath, bm}
\usepackage{graphicx}
\usepackage{physics}
\usepackage[colorlinks=true, allcolors=blue]{hyperref}

\setcitestyle{super}

\renewcommand{\bra}[1]{\left\langle#1\right|}
\renewcommand{\ket}[1]{\left|#1\right\rangle}
\renewcommand{\vec}[1]{\boldsymbol{#1}}

\begin{document}

\title{Flat-band ratio and quantum metric in the superconductivity of modified Lieb lattices}
\author{Reko P. S. Penttilä}
\email{reko.penttila@aalto.fi}
\affiliation{Department of Applied Physics, Aalto University School of Science, FI-00076, Espoo, Finland}

\author{Kukka-Emilia Huhtinen}
\affiliation{Department of Applied Physics, Aalto University School of Science, FI-00076, Espoo, Finland}
\affiliation{Institute for Theoretical Physics, ETH Zurich, CH-8093, Zurich, Switzerland}

\author{Päivi Törmä}
\email{paivi.torma@aalto.fi}
\affiliation{Department of Applied Physics, Aalto University School of Science, FI-00076, Espoo, Finland}

\begin{abstract}
Flat bands may offer a route to high critical temperatures of superconductivity. It has been predicted that the quantum geometry of the bands as well as the ratio of the number of flat bands to the number of orbitals determine flat band superconductivity. However, such results have assumed at least one of the following: an isolated flat band, zero temperature, mean-field theory, and/or uniform pairing. Here, we explore flat band superconductivity when these assumptions are relaxed. We consider an attractive Hubbard model for different extensions of the Lieb lattice. The superconducting order parameter, critical temperature, superfluid weight, and Berezinskii-Kosterlitz-Thouless (BKT) temperature are calculated within dynamical mean-field theory. We find that for isolated flat bands, the flat-band ratio and quantum geometry are in general good indicators of superconductivity even at finite temperatures. For non-isolated flat bands, a good guideline of the BKT temperature is provided by the zero-temperature superfluid weight and the flat-band ratio.

\end{abstract}

\maketitle
\section{Introduction}
Both theory and experiments suggest that flat bands can be beneficial for superconductivity. Studies of superconductivity on twisted bilayer graphene, a material with (nearly) flat bands at certain twist angles, have shown one of the highest critical temperature ($T_c$) to charge carrier density ratios ever measured~\cite{Pablo2018,MacDonald2019,Andrei2020TBG,Andrei2021,Efetov2024}. On the theory side, flat bands have been studied in the repulsive Hubbard model for instance in the context of ferromagnetism~\cite{Lieb1989,Mielke1991,Tasaki1992,Mielke1993,Tasaki1998} and superconductivity~\cite{Kuroki2005,Kiesel2012,Aoki2020}.  This work focuses on attractive interactions, where even the simple Bardeen-Cooper-Schrieffer (BCS) theory predicts that the critical temperature for flat bands is proportional to the effective attractive interaction $|U|$\cite{heikkila2011,shaginyan1990superfluidity}, while the critical temperature for dispersive bands is given by $T_c \propto \mathrm{exp}\left(-1/|U|\rho_0\right)$, where $\rho_0$ is the density of states at the Fermi surface. This indicates that $T_c$ could be greatly enhanced in flat bands.

Furthermore, it has been theoretically shown that flat bands can have a non-zero superfluid weight and thus support dissipationless transport~\cite{peotta2015}, which is not immediately clear since non-interacting flat-band states are localized. In an isolated perfectly flat band, the superfluid weight (stiffness) $D_s$ is completely of quantum geometric origin, i.e.~a non-zero quantum metric~\cite{Provost1980} enables supercurrent~\cite{peotta2015,Rossi2021,Torma2022,peotta2023quantum}. Since the quantum metric is bounded by topological invariants such as Chern numbers and winding numbers~\cite{Roy2014,Ozawa2021,Mera2022}, nontrivial topology can enable flat band superconductivity~\cite{peotta2015,Xie2020}. At zero temperature, for flat bands isolated from other bands by a gap larger than the interaction scale $|U|$ ("isolated flat bands") and fulfilling the so-called uniform pairing condition~\cite{tovmasyan2016} which requires the order parameters to be equal at all orbitals where the flat band states reside, the superfluid weight is proportional to the minimal quantum metric~\cite{huhtinen2022,peotta2015} and to the ratio of the number of degenerate flat bands to orbitals where the flat band states reside ("flat-band ratio")~\cite{herzog2022}. These results have been derived both within mean-field theory~\cite{huhtinen2022,peotta2015} and by exact calculations at zero temperature~\cite{herzog2022,huhtinen2022} --- always assuming the isolated flat band and uniform pairing conditions.

In the search of new superconducting materials with high critical temperatures and other desirable properties such as high critical currents, efficient rules of thumb as well as relations between superconductivity and simple band structure properties (quantum geometry, number of flat bands), would be immensely useful. It is therefore of interest to explore whether the existing simple relations hold beyond the restrictive conditions for which they have been derived. The analytical prediction that the flat band superfluid weight and critical temperature is linearly proportional to the interaction, $D_s \propto |U|$, derived by mean-field for isolated flat bands~\cite{peotta2015}, has been verified by various numerical methods including quantum Monte Carlo, density matrix renormalization group and dynamical mean-field (DMFT) calculations~\cite{Julku2016,Peri2021,Herzog-Arbeitman2022,Chan2022,Chan2022b,Hofmann2023,Hofmann2020,Orso2022}. There exist also a few studies on how to describe the case of flat bands touching other bands~\cite{Iskin2019,Iskin2022,herzogarbeitman2024}, within mean-field or effective theories. For some lattice models, expressions have also been obtained for the zero-temperature mean-field superfluid weight in flat bands not fulfilling the uniform pairing condition~\cite{Chan2022}. To the best of our knowledge, systematic finite-temperature beyond-mean-field studies of the validity of the relations between the superfluid weight and the quantum metric, as well as the flat-band ratio, are missing. The goal of this article is to take the first steps in this direction.   

We chose as our model system various extensions of the Lieb lattice, since this will allow us to systematically change the number of flat bands and the quantum geometry. The calculations of the superconducting order parameter and the superfluid weight are done using DMFT, as it can describe the local (in our case within the unit cell) quantum and thermal fluctuations exactly. However, DMFT becomes exact only in the limit of large coordination number, which in our lattices varies between one and six. We show by comparison to a mean-field calculation that such a beyond mean-field approach is necessary. 

In the "System" subsection of the Results section, we present the different lattice models studied in this work and provide the Hamiltonian used. In the Results section, we present superconducting order parameters and superfluid weights for the considered lattices calculated with DMFT. We show that the quantum metric and the flat-band ratio can be good predictors of superconductivity at zero or very low temperatures, even if the system is not in the isolated flat band and uniform pairing limit. Furthermore, they can even predict the critical temperatures of superconductivity in most of the lattices considered. Our results point to a potential method to design flat band systems with increased critical temperatures: the addition of orbitals resulting in an increased number of flat bands at the Fermi energy.

\begin{figure*}
	\centering
	\includegraphics{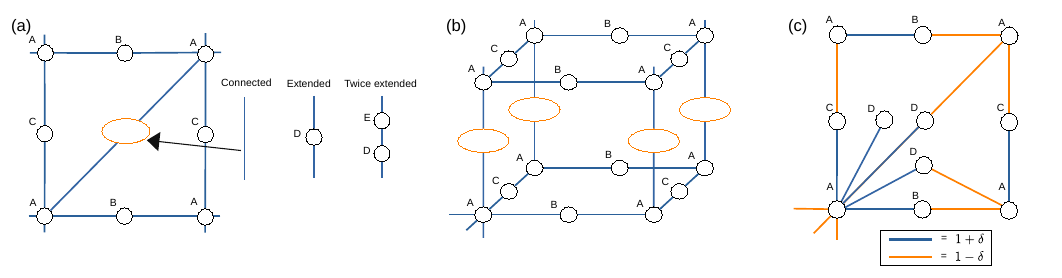}
	\caption{\textbf{Considered lattices.} a) Different two-dimensional extensions of the Lieb lattice called connected, extended, and twice-extended, which are obtained by inserting either a hopping connection or one or two additional orbitals in the place of the orange ellipse. The distinct orbitals in the unit cell are denoted with $A$, $B$, $C$, $D$ and $E$. b) Different three-dimensional extensions obtained by inserting the extensions in the places of the orange ellipses. c) Different versions of the two-dimensional extended Lieb lattice called diagonal, x-directional and decorated. The hopping staggering parameter $\delta$  is used to modify the hoppings inside and between unit cells.}
	\label{fig:lattices}
\end{figure*}

\section{Results}
\label{sec:results}

\subsection{The system}
\label{sec:system}
In this study, we investigate different extensions of the Lieb lattice in two and three dimensions. The two-dimensional Lieb lattice is a bipartite lattice (see Fig.~\ref{fig:lattices} a) without the diagonal line from site $A$ to another site $A$) featuring a single zero-energy flat band. In the repulsive Hubbard model, it has a ferromagnetic ground state at half-filling~\cite{Lieb1989}. Through a particle-hole transformation, the repulsive Hubbard model can be mapped to an attractive one. This transformation maps the ferromagnetic wavefunction to a BCS wavefunction~\cite{Julku2016}. 

The extensions of the Lieb lattice allow us to change the number of flat bands and tune the quantum metric of the lattice. Fig.~\ref{fig:lattices} a) shows three different extensions in two dimensions, all constructed by connecting the next nearest $A$ sites with either a straight hopping term or additional orbitals. We name the different extensions in the following way: the straight hopping connection is the connected Lieb lattice, one additional orbital is the extended Lieb lattice and the extension with two additional orbitals is the twice-extended Lieb lattice. The blue lines give the allowed hoppings with a hopping energy of $t=1$. All energies and temperatures in this article are given in units of the hopping $t$, and we set the Planck's and Boltzmann constants and the electron charge to unity, $\hbar=k_B=e=1$. 

Fig.~\ref{fig:lattices} b) shows different three-dimensional extensions of the Lieb lattice, which are constructed by stacking two-dimensional Lieb lattices and connecting them from the $A$ sites with either a direct hopping connection or with additional orbitals. The three-dimensional extensions are named similarly to the two-dimensional extensions. All of these extensions in both two and three dimensions preserve the original zero-energy flat band of the Lieb lattice since the localized flat-band states in the Lieb lattice reside only on the $B/C$ orbitals and we make the extensions from the $A$ orbitals. In the Lieb, connected and twice extended lattices, there is a single flat band with states residing at the $B/C$ orbitals. In the extended Lieb lattice, there are two flat bands with states residing at the $B/C/D$ orbitals. The ratio of the number of degenerate flat bands, $N_f$, to the number of orbitals at which the flat band states reside, $N_{of}$, called the flat-band ratio, is therefore different: in the Lieb, connected and twice extended lattices $N_f/N_{of} = 1/2$, whereas in the extended Lieb lattices $N_f/N_{of} = 2/3$.

Fig.~\ref{fig:lattices} c) presents different versions of the two-dimensional extended Lieb lattice, called the diagonal, x-directional, and decorated versions. In the diagonal version, the extension is made diagonally from the $A$ site to the $A$ site of the next-nearest unit cell, while in the x-directional version, the extension is made to the next unit cell in the x-direction. In the decorated Lieb lattice, an additional orbital is added to the unit cell and connected only to the $A$ orbital of the same unit cell. All of these different versions have two flat bands with states residing at three different orbitals, resulting in a flat-band ratio of $N_{f}/N_{of} = 2/3$. The hoppings are altered with a hopping staggering parameter $\delta$, which increases the hoppings inside the unit cell (blue lines) and decreases the hoppings between unit cells (orange lines). Setting $\delta\neq 0$ will allow us to open a band gap in those lattices that have a Dirac cone touching the flat band, which makes the integrated quantum metric of the flat band, used in our analysis, finite.

We study the multi-band Hubbard Hamiltonian with an attractive onsite interaction $U<0$
\begin{equation}
    H = \sum_{<n,n'>} t_{n,n'} c^\dagger_{ n\sigma } c_{n'\sigma} + \mu \sum_n c^\dagger_{n\sigma }c_{n\sigma } + U \sum_{n}c^\dagger_{ n\uparrow} c_{n\uparrow} c^\dagger_{ n\downarrow} c_{n\downarrow}, \label{eq.hubbard}
\end{equation}
where $n \equiv i, \alpha$ labels the unit cell $i$ and the orbital $\alpha$. The solutions of this Hamiltonian are given by Green's functions, with the Nambu-Gorkov formalism where, for a lattice with $M$ orbitals in the unit cell, both the Green's functions and self-energies are $2M \times 2M$ matrices. This is essential since we have to include anomalous components that describe Cooper pairing in the system. The Green's functions and the self-energy are then given by
\begin{align}
    \bm{G}_i(i\omega_n) &= \begin{bmatrix} G_i(i\omega_n) &  F_i(i\omega_n)\\
    F_i^*(i\omega_n) & -G_i^*(i\omega_n)
    \end{bmatrix},\\
    \bm{\Sigma_i}(i\omega_n) &= \begin{bmatrix}\Sigma_i(i\omega_n) &  S_i(i\omega_n)\\
    S_i^*(i\omega_n) & -\Sigma_i^*(i\omega_n)
    \end{bmatrix}, 
    \label{eq:Nambu Green's and self energy}
\end{align}
where $G_i(i\omega_n)$ and $\Sigma_i(i\omega_n)$ ($F_i(i\omega_n)$ and $S_i(i\omega_n)$) are the normal (anomalous) components of the Green's function and the self-energy, respectively. All of these components are $M\times M$ matrices.
We assume the Green's functions and self-energies depend only on the relative unit cell index $i$ because of the translational invariance of the lattice. Because we use real-space DMFT, the self-energy is approximated to be diagonal in the orbital indices (see section Methods). The order parameters of each orbital can be obtained from the diagonal elements of the anomalous components of the local Green's functions, $F_{loc}(i\omega_n)$,  which we calculate using dynamical mean-field theory (DMFT):
\begin{equation}
    \Delta_\alpha = \left[ F_{loc}(i\omega_n)  \right]_{\alpha, \alpha}.
    \label{eq:order parameter}
\end{equation}
We also calculate the superfluid weight for the presented lattices with DMFT using linear response theory (details are presented in the Methods section).

\begin{figure*}[t]
	\centering
	\includegraphics{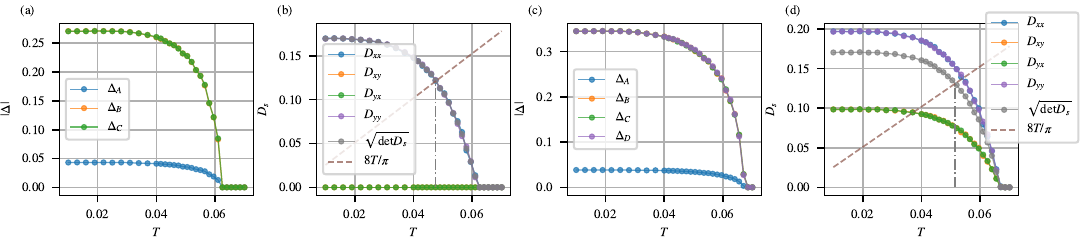}
	\caption{\textbf{Temperature dependence of the order parameters and superfluid weights for the Lieb and the twice-extended Lieb lattices.} Absolute values of order parameters as a function of temperature for (a) the original Lieb lattice and (c) the two-dimensional extended Lieb. The order parameters of orbitals $B$ and $C$ are equal in a) and thus visible as one line only. Same holds true for orbitals $B$, $C$ and $D$ in c). The components of the superfluid weights and $\sqrt{\mathrm{det} \,D_s}$ for (b) the original Lieb lattice  and (d) the two-dimensional extended Lieb lattice. Both the diagonal components $D_{xx}$ and $D_{yy}$ are equal in the original Lieb and the extended Lieb lattices, and $D_{xy}=D_{yx}$. The dashed brown line is given by $8T/\pi$ and is used to determine the $T_{BKT}$ which is shown by the dashed gray line.}
	\label{fig:fig2}
\end{figure*}

\begin{figure*}
	\centering
	\includegraphics{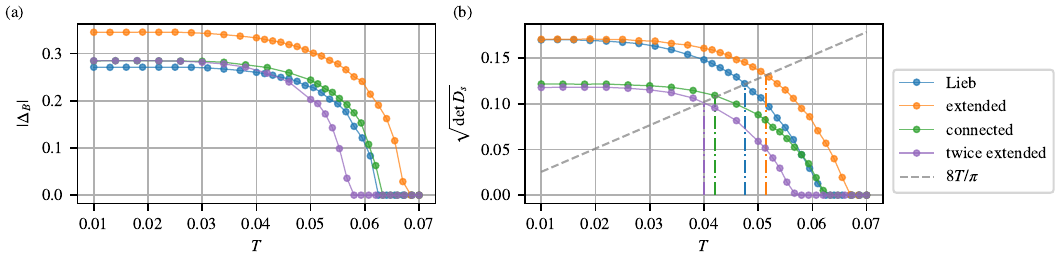}
	\caption{\textbf{Temperature dependence of the order parameters and superfluid weights for the different two-dimensional extensions.} a) Absolute value of the order parameter $|\Delta_B|$ as a function of temperature for the original Lieb and the two-dimensional extensions (see Fig.~\ref{fig:lattices} a)), with interaction strength $|U|=1$. b) The square root of the determinant of the superfluid weight for the original Lieb lattice  and the different two-dimensional extensions, with $|U| = 1$. The dashed gray line is given by $8T/\pi$ and is used to determine the $T_{BKT}$, which is shown for each lattice by the dashed-dotted lines.}
	\label{fig:fig3}
\end{figure*}

\subsection{Two-dimensional extensions} \label{sec.2d_extensions}
In this section, we present the results of DMFT calculations for the different two-dimensional extensions shown in Fig.~\ref{fig:lattices} a). The quantities of interest are the order parameters and superfluid weight as a function of temperature. Figures~\ref{fig:fig2} a) and c) show the order parameters of the original Lieb lattice and two-dimensional extended Lieb lattice, respectively. The  interaction strength is $|U|=1$ and the chemical potential is $\mu=0$, corresponding to a half-filled flat band in both of the lattices. We choose to work at half-filling because mean-field theory predicts that it maximizes the superfluid weight~\cite{peotta2015,Julku2016}. We see that the critical temperatures of the order parameters, i.e., the temperature at which the order parameters vanish, increase from the original to the extended Lieb lattice. Moreover, the absolute values of the order parameters are larger for the extended Lieb geometry. 

Figures~\ref{fig:fig2} b) and d) show all the components of the superfluid weight and the square root of its determinant for the original and extended Lieb lattices, respectively.  The superfluid weights become zero at the critical temperature of the order parameters of the corresponding lattices. However, the real critical temperature is given by the BKT transition temperature, which is $T_{BKT}\approx0.051$ for the extended Lieb lattice and $T_{BKT}\approx0.047$ for the original Lieb lattice. These results indicate that this extension of the Lieb lattice benefits superconductivity because of the larger $T_{BKT}$.

Next, we present the results of the same calculation for all two-dimensional extensions shown in Fig.~\ref{fig:lattices} a). Figure~\ref{fig:fig3} a) shows $|\Delta_B|$ for these extensions and the original Lieb lattice. It is enough to look only at $|\Delta_B|$ since, even though the order parameters do not have the same amplitude at all orbitals, they vanish at the same temperature. These results are obtained from DMFT calculations with attractive interaction strength $|U|=1$ and chemical potential is chosen so that the lattices are half-filled. We see that while the critical temperature $T_C$ of $|\Delta_B|$ for the extended Lieb lattice is increased from the original Lieb lattice, $T_C$ stays approximately the same for the connected Lieb lattice. 

These findings are qualitatively in line with the mean-field result $T_C \approx |U| N_f/(4 N_{of})$, which shows that the critical temperature for the order parameters is approximately proportional to the flat-band ratio $N_f/N_{of}$. This result has been obtained assuming the uniform pairing condition and isolated flat bands (see Supplementary Note 3). In the Lieb and connected lattices $N_f/N_{of}=1/2$ and $T_C$ is approximately equal while in the extended Lieb lattice $N_f/N_{of}=2/3$ and $T_C$ is larger. However, in the twice-extended Lieb lattice, the flat-band ratio is also $1/2$, but $T_C$ is less than in the Lieb and connected lattices. This difference is not present in mean-field results, shown in Supplementary Note 4. This shows that the estimate $T_C \approx |U| N_f/(4 N_{of})$ does not capture the full behavior of the critical temperature.

We note here that this mean-field result is similar to a lower bound for general (uniform pairing condition not assumed) bipartite lattices $|\left<\Delta^B \right>| \geq |U| N_f/(2 N_{of})$, where $|\left<\Delta^B \right>|$ is the average of the order parameters of the larger sublattice at zero temperature~\cite{bouzerar2024}. We also confirm that for all of the bipartite lattices studied $|\left<\Delta^B \right>|$ is indeed larger than the lower bound. While the lower bound is derived for bipartite lattices with half filled flat bands, we find that even for non-bipartite lattices (the stacked and twice extended two- and three-dimensional Lieb lattices) $|\left<\Delta^B \right>|$ is larger than the lower bound. Furthermore, the largest 
\begin{equation}
    \frac{|\left<\Delta^B \right>| -|U| N_f/(2 N_{of})}{|\left<\Delta^B \right>|}
\end{equation}
in bipartite lattices is 0.077 (Lieb lattice) and in non-bipartite lattices 0.12 (twice extended three-dimensional Lieb), which shows that the lower bound is close to $|\left<\Delta^B \right>|$.

Figure~\ref{fig:fig3} b) shows the square roots of the determinants of the superfluid weights and the BKT transition temperatures for the two-dimensional extensions. The superfluid weights vanish at the same temperatures as the order parameters, and thus, these critical temperatures of the superfluid weights also mirror the behavior of the flat-band ratio for the connected, extended and Lieb lattices. However, the BKT transition temperatures behave qualitatively differently. The original Lieb lattice has a larger $T_{BKT}$ than the connected Lieb lattice, even though these have the same flat-band ratio. This shows that while the flat-band ratio succeeds at explaining the behavior of $T_c$ for the Lieb, connected and extended lattices, it fails to fully explain the behavior of the BKT transition temperature.

In Supplementary Note 4), we show the mean-field comparison of these superfluid weight results (see Supplementary Figure S4 a). At the mean-field level, the zero temperature $\sqrt{\det D}$ behaves qualitatively similar to the DMFT results, but is slightly larger in all lattices. The mean-field critical temperatures are approximately 2.5 times higher, with distinct qualitative differences: the twice-extended lattice has approximately the same critical temperature as the original Lieb and the connected lattices, which all have the same flat-band ratio, while the $T_c$ of the extended lattice is around 1.3 times larger. Furthermore, we see that the mean-field $T_c$ of all lattices is close to the isolated flat-band result $T_c \approx |U| N_f/(4 N_{of})$, which shows that at the mean-field level, the flat-band ratio can predict $T_c$ in all of the lattices, including the twice-extended one. However, also at the mean-field level, we see that the flat-band ratio cannot predict the BKT transition temperatures.

\begin{figure*}
	\centering
	\includegraphics{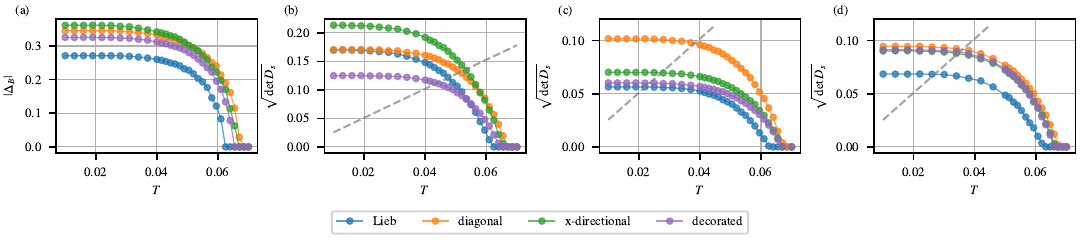}
	\caption{\textbf{Order parameters and superfluid weights for the different versions of the extended Lieb lattice with different hoppings staggering $\delta$.} a-b) The absolute value of the order parameter $|\Delta_B|$ and the square root of the determinant of the superfluid weight as a function of temperature for the different versions of the extended Lieb lattice with the  hopping staggering $\delta = 0$. c) The square root of the determinant of the superfluid weight as a function of temperature for different versions of the extended Lieb with $\delta = 0.3$. d) The square root of the determinant of the superfluid weight as a function of temperature for different versions of the extended Lieb with the same determinant of the minimal quantum metric $\det M = 1.3$. Results in all panels have been calculated with the interaction strength $|U| = 1$.}
	\label{fig:fig4}
\end{figure*}

\begin{figure}
	\centering
	\includegraphics{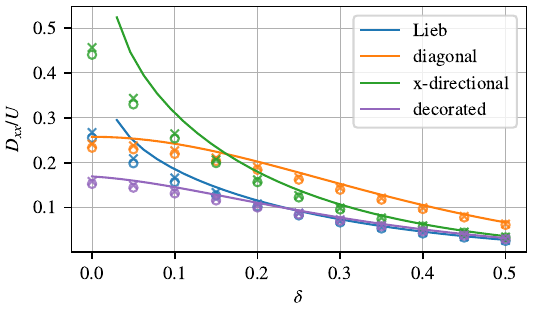}
	\caption{\textbf{Comparison of zero-temperature superfluid weights obtained from the quantum metric prediction and DMFT and mean-field calculations.} The $D_{xx}$ component of the superfluid weight as a function of the hopping staggering $\delta$ at the temperature $T=0.003$ with the interaction strength $|U|=1/3$ for the original Lieb and the different versions of the two-dimensional extended Lieb. The points give the DMFT $D_{xx}$, and the crosses give $D_{xx}$ obtained from mean-field calculations including Hartree-terms, and the lines give the predictions calculated using Eq.~\eqref{eq.ds_qm}.}
	\label{fig:Dxx_pred}
\end{figure}

\subsection{Influence of the quantum metric}
We also study different versions of the extended Lieb lattice shown in Fig.~\ref{fig:lattices} c). These different versions are called the diagonal, the x-directional and the decorated versions. All of these lattices have two degenerate flat bands at zero energy and three orbitals where these flat bands reside, resulting in a flat-band ratio of $2/3$, which is larger than the $1/2$ ratio of the original Lieb lattice. Thus, based on the results from the previous section, we could expect the critical temperatures of all the different versions to be larger than that of the original Lieb lattice. Figure~\ref{fig:fig4} a) confirms this expectation, showing higher critical temperatures across the different versions compared to the original Lieb lattice. However, while the flat-band ratio predicts that $T_c$ of the different versions is approximately 1.3 times that of the original lattice, the DMFT critical temperatures are much closer to the original one. 

Figure~\ref{fig:fig4} b) shows the square roots of the determinants of the superfluid weights $\sqrt{\mathrm{det} D}$ and the BKT transition temperatures $T_{BKT}$ for the different versions. The flat-band ratio does not predict the behavior of $T_{BKT}$ since all the different versions have different BKT temperatures while having the same flat-band ratio. Furthermore, $T_{BKT}$ is larger for the original Lieb lattice than the decorated version, even though the decorated version has a larger flat-band ratio. These results show that while the flat-band ratio can predict the qualitative behavior of the critical temperature of the superfluid weight, it fails to predict $T_{BKT}$, which is the true critical temperature of superconductivity in 2D. However, we see that the zero temperature $\sqrt{\mathrm{det} D}$, which usually gives only the upper limit for the $T_{BKT}$, can predict the qualitative behavior of $T_{BKT}$ in the different versions of the extended Lieb lattice: x-directional version has the largest $T_{BKT}$ followed by the diagonal and then the decorated version, exactly in the order of their zero temperature $\sqrt{\mathrm{det} D}$. Furthermore, we see that $T_{BKT}$ of the original Lieb is smaller than that of the extended Lieb, even though these have the same zero temperature $\sqrt{\mathrm{det} D}$. This is caused by the smaller $T_c$ of the original Lieb, which is, in turn, predicted by the smaller flat-band ratio. Thus, the flat-band ratio and the zero-temperature superfluid weight combined provide a good estimate of the behavior of $T_{BKT}$.

At the mean-field level, the zero-temperature superfluid weight of isolated flat-band lattices can be predicted by the minimal quantum metric and the flat-band ratio. The quantum metric of a certain set of bands $S$ is the real part of the quantum geometric tensor
\begin{equation}
    \bm{\beta}_{ij}(\bm{k}) = 2 \mathrm{Tr}P(\bm{k})\partial_i P(\bm{k})\partial_j P(\bm{k}),
\end{equation}
where $P(\bm{k}) = \sum_{\beta\in S}\ket{\beta_{\bm{k}}}\bra{\beta_{\bm{k}}}$ is the projector of the eigenstates to the Bloch states of the bands in the set $S$. In the isolated flat-band limit the square root of the determinant of the zero temperature superfluid weight is proportional to the minimal quantum metric $M^{\rm min}$~\cite{huhtinen2022, herzog2022}:
\begin{equation}
    \sqrt{\mathrm{det} D_s} = \frac{4f(1-f)}{(2 \pi)^{D-1}}\frac{N_f}{N_{of}}|U|\sqrt{\mathrm{det}M^{\rm min}}, \label{eq.ds_qm}
\end{equation}
where $D$ is the dimension, $f$ is the filling factor of the flat bands, $U$ the attractive interaction strength, and $N_{of}$ ($N_f$) is the number of orbitals where the flat band states reside (degenerate flat bands). Here $\mathrm{det}M^{\rm min}$ denotes the determinant of the minimal quantum metric which is obtained by minimizing the trace of the integrated quantum metric \begin{equation}
    M_{ij} = \frac{1}{2\pi}\int_{\mathrm{B.Z.}}d^2\bm{k} \; \mathrm{Re}(\bm{\beta}_{ij}(\bm{k})),
\end{equation}
with reference to the positions of the orbitals in the unit cell, while tight-binding parameters are kept constant. In addition to the isolated flat-band limit, Eq.~\eqref{eq.ds_qm} assumes the presence of time-reversal symmetry and uniform pairing, i.e. that order parameters are equal in those orbitals where the flat band states reside.

In Fig.~\ref{fig:Dxx_pred}, we compare the zero temperature superfluid weights as a function of the hopping staggering $\delta$, calculated with DMFT, mean-field theory including Hartree-terms, and with Eq.~\eqref{eq.ds_qm}. The lattices are at half-filling and the DMFT and mean-field calculations were conducted with $|U|=1/3$ and $T=0.003$ (which gives results very close to those at zero temperature). The hopping staggering $\delta$ increases the intra-unit-cell hoppings and decreases the hoppings between unit cells, which opens the band gap. For the diagonal and decorated versions, the superfluid weight from Eq.~\eqref{eq.ds_qm} matches the DMFT one extremely well, while the prediction fails for the x-directional version and original Lieb lattice with small $\delta$. This difference stems from the band structures shown in Supplementary Note 1: the Lieb and the x-directional lattices have singular band touchings with $\delta=0$~\cite{Rhim2021}, for which the quantum metric diverges, while the flat bands of the diagonal and decorated lattices are always separated. From Fig.~\ref{fig:Dxx_pred}, we also see that, in the small interaction regime, the superfluid weight obtained from the mean-field calculations is almost precisely equal to the DMFT one in all of the lattices. 

In addition to the isolated flat band condition, Eq.~\eqref{eq.ds_qm} is derived assuming uniform pairing. From Fig.~\ref{fig:fig2} c), we see that $|\Delta_A|$ is different from the other order parameters for the diagonal version of the extended Lieb lattice. This is, in fact, true for all the different extensions and the original Lieb lattice. The flat-band states reside on the $B$, $C$ and $D$ orbitals, which is why the different $\Delta_A$ does not affect the prediction much: we need uniform pairing at the orbitals where the flat-band states reside, whereas the order parameters at the other orbitals should be small in the isolated flat band limit. In the decorated and the x-directional lattices, even the order parameters of the $B$, $C$ and $D$ orbitals are not equal:$|\Delta_C|$ differs from $|\Delta_B|$ and $|\Delta_D|$ in the x-directional version and $|\Delta_D|$ differs from $|\Delta_B|$ and $|\Delta_C|$ in the decorated version. Despite these differences in the order parameters, the Eq.~\eqref{eq.ds_qm} can predict the DMFT (and mean-field) superfluid weigh when the band gap is sufficiently large. This indicates that the prediction works also when the uniform pairing condition is not perfectly met.

 With a larger interaction strength $|U| = 1$, both the Eq.~\eqref{eq.ds_qm} and the mean-field calculation give larger zero-temperature superfluid weights compared to DMFT (see Supplementary Note 2). However, in all of the lattices, the differences between the superfluid weights are only quantitative and the superfluid weights have the same behavior as a function of $\delta$. Thus, while DMFT agrees quantitatively with Eq.~\eqref{eq.ds_qm} and the mean-field calculation only at small interaction strength, there is still qualitative agreement at higher interaction strengths. Specifically, if two lattices exhibit distinct DMFT zero-temperature superfluid weights, the ratio of these superfluid weights should be approximately similar in the mean-field calculations or in the Eq.~\eqref{eq.ds_qm} predictions when the flat bands are sufficiently isolated. These results show that when the bands are isolated, mean-field theory can be used instead of the DMFT zero-temperature superfluid weight when comparing the relative BKT temperatures of different lattices.

Fig.~\ref{fig:fig4} c) shows  $\sqrt{\det D}$ for the different versions of the extended and the original Lieb lattice with a hopping staggering parameter $\delta = 0.3$. The critical temperatures of $\sqrt{\det D}$ are almost identical to the $\delta=0$ ones, with only minor deviations: the decorated and x-directional versions exhibit slightly higher $T_c$ with $\delta=0.3$. Now, since all the lattices have band gaps, the zero-temperature $\sqrt{\det D}$ obtained from Eq.~\eqref{eq.ds_qm} and shown in Table~\ref{tab:quantum_metrics} behave qualitatively similarly as the DMFT results~(Fig.~\ref{fig:fig4} c extrapolated to zero temperature). Specifically, the diagonal version exhibits the highest zero temperature superfluid weight, followed by the x-directional version, while the original Lieb and decorated versions have nearly identical values. Furthermore, we see that the qualitative differences in the BKT temperatures between the different extensions are similar to the differences of the zero-temperature superfluid weights obtained from Eq.~\eqref{eq.ds_qm}. Therefore, the flat-band ratio and the quantum metric predict $T_{BKT}$ reasonably well.

Fig.~\ref{fig:fig4} d) shows $\sqrt{\mathrm{det} D}$ for the different versions with such $\delta$ parameters that all the lattices have the same determinant of the minimal quantum metric of 1.3. The zero-temperature $\sqrt{\mathrm{det} D}$ obtained from Eq.~\eqref{eq.ds_qm} is $0.121$ for all three versions of the extended Lieb lattice and $0.091$ for the original Lieb because of the smaller flat-band ratio. As in the $\delta = 0.3$ case, these predictions overestimate the DMFT results but accurately predict the qualitative behavior: all different versions have almost equal $\sqrt{\mathrm{det} D}$ at zero temperature, which is larger than that of the original Lieb lattice. Furthermore, we find that all the versions of the extended Lieb lattice have approximately identical $T_{BKT}$, which is larger than that of the original Lieb as predicted by the minimal quantum metric and the flat-band ratio. 

The results in this section demonstrate that the flat-band ratio and the minimal quantum metric are effective in predicting the qualitative behavior of $T_{BKT}$ across different lattices: lattices with larger flat-band ratios and minimal quantum metrics correspond to higher $T_{BKT}$ values. This approach is valid, however, only when the flat bands are sufficiently isolated. In cases where they are not, the mean-field zero-temperature superfluid weight can be used in place of the Eq.\eqref{eq.ds_qm}, in combination with the flat-band ratio, to predict the BKT temperature. In Supplementary Note 4, we have performed mean-field calculations for the systems considered in Figs.~\ref{fig:fig4} a) and b) and compared them to the DMFT results. While at the mean-field level, the critical temperatures and $T_{BKT}$ are larger, they have the same qualitative behavior as in the DMFT results, i.e, $T_{BKT}$ can be predicted from the zero-temperature superfluid weight and the flat-band ratio, or the flat-band ratio and the minimal quantum metric in flat bands that are gapped.

\begin{table}
\caption{\textbf{Zero-temperature superfluid weights calculated from the quantum metric prediction.} Predicted zero-temperature superfluid weights for the different versions of the extended Lieb with the hopping staggering $\delta = 0.3$. Here $\sqrt{\text{det} D(0)}$ is calculated using Eq.~\eqref{eq.ds_qm}.}
\label{tab:quantum_metrics}
\begin{center}
\begin{tabular}{ |c||c|c|} 
 \hline
 Lattice  & $\sqrt{\mathrm{det} D(0)}$ from quantum metric \\ 
 \hline
 original Lieb &  0.073\\ 
 Diagonal extended & 0.130  \\
 X-directional extended & 0.091\\
 Decorated & 0.077\\
 \hline
\end{tabular}
\end{center}
\end{table}

\begin{figure}
	\centering
	\includegraphics{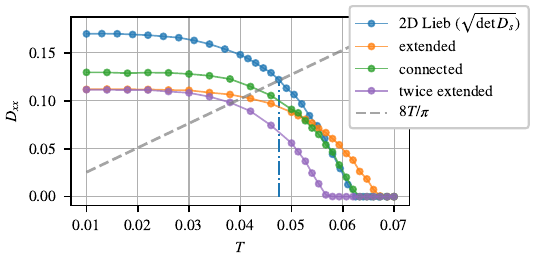}
	\caption{\textbf{Order parameters and superfluid weights for the three-dimensional extensions and the original Lieb lattice.} $D_{xx}$ components of the superfluid weights of the different three-dimensional extensions and $\sqrt{\mathrm{det} D_s}$ of the two-dimensional Lieb lattice as a function of temperature. The dashed gray line is given by $8T/\pi$ and is used to determine the $T_{BKT}$ for the original 2D Lieb lattice, which is shown by the dashed-dotted blue line.}
	\label{fig:Lieb_3D}
\end{figure}

\subsection{Three-dimensional extensions}
Fig.~\ref{fig:lattices} shows how the two-dimensional Lieb lattice can be extended to three dimensions in a similar way as in Fig.~\ref{fig:lattices}, a), i.e., the different extensions are made from the smaller sublattice. This preserves the flat band of the two-dimensional Lieb lattice. Once again, the (twice-) extended three-dimensional Lieb lattice has (two) one additional orbital(s) between the 2D planes. In the connected three-dimensional Lieb lattice, there is a straight hopping connection between the $A$ sites. The flat-band ratios are the same in the three-dimensional extensions as in the two-dimensional ones, i.e., $2/3$ in the extended, whereas the twice-extended and connected lattices have the same flat-band ratio of $1/2$ as the Lieb lattice. 

Fig.~\ref{fig:Lieb_3D} shows $D_{xx}$ for the different three-dimensional extensions of the lattice and $\sqrt{\mathrm{det}D}$ for the original two-dimensional Lieb lattice. The temperatures at which the superfluid weight vanishes for the three-dimensional extensions follow the flat-band ratio similarly as in the two-dimensional extensions, i.e., the extended lattice has the largest critical temperature while connected has the second largest and twice extended the lowest. The temperature at which $\sqrt{\mathrm{det}D}$ vanishes for the two-dimensional Lieb equals the critical temperature of the connected extension. However, in the two-dimensional lattice, the real critical temperature of the superfluid weight is given by the BKT-transition temperature shown in Fig.~\ref{fig:Lieb_3D}. In three dimensions, long-range order exists, and the real critical temperature of the superfluid weight is given by the temperature at which the superfluid weight vanishes. Thus, the critical temperature of superconductivity increases for all three-dimensional extensions compared to the two-dimensional Lieb lattice. 

For these models, DMFT results can differ significantly from mean-field predictions, both qualitatively and quantitatively, especially for small interlayer hopping amplitudes (see Supplementary Note 4). For small $d$, mean-field theory can significantly overestimate $T_c$ in the 3D extended Lieb lattice because it predicts a large nonzero order parameter at the weakly connected site $D$ between layers, contrary to DMFT. 

\subsection{Qualitative behavior of the superfluid weight and order parameters}

\begin{figure*}
	\centering
	\includegraphics{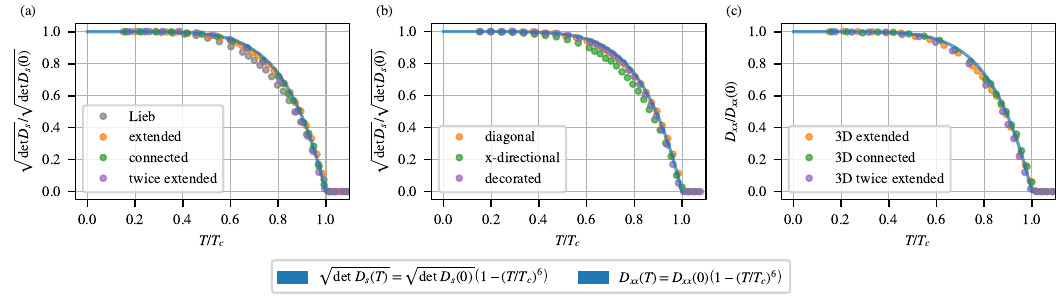}
	\caption{\textbf{Qualitative temperature behavior of the superfluid weights.} a) and b) Determinant of the superfluid weight scaled by the critical temperature and the zero temperature superfluid weight for all the two-dimensional modified Lieb lattices with the interaction strength $|U|=1$. c) $D_{xx}$ component of the superfluid weight scaled by the critical temperature and the zero temperature superfluid weight for the three-dimensional extensions with $|U|=1$.}
	\label{fig:scaled_sfw}
\end{figure*}

Figure~\ref{fig:scaled_sfw} shows that the superfluid weights of all the studied lattices have approximately the same behavior as a function of temperature. In the two-dimensional cases, the temperature behavior of ${\rm det} D_s$ agrees with $\sqrt{\mathrm{det} D(T)} = \sqrt{\mathrm{det} D(0)}(1-(T/T_c)^b)$, with $b \approx 6$ for all lattices expect the x-directional and original Lieb, for which $b$ is smaller and closer to 5. For the three dimensional lattices Fig.~\ref{fig:scaled_sfw} c) shows that the qualitative behavior of $D_{xx}$ is in agreement with $ D_{xx}(T) = D_{xx}(0)(1-(T/T_c)^6)$. This behavior is distinct from the temperature behavior obtained from mean-field calculations where $b \approx 3$ matches the temperature behavior of the superfluid weights better. 

We also find that in lattices with uniform pairing the order parameters of the orbitals in which the flat-band states reside agree with $\Delta(T) = \Delta(0)\sqrt{1-(T/T_c)^6}$ (see Supplementary Note 5). Thus, in lattices with UPC, the temperature behavior of the superfluid weight is approximately the same as the behavior of $\Delta(T)^2$. We find that in the original Lieb lattice the temperature behavior of this relation does not hold equally well since the obtained $b$ for superfluid weights is closer to 5. This relation is also present in Eq.~\eqref{eq.ds_qm} since in isolated flat bands with UPC the order parameter is given by $\Delta = N_f/N_{of} |U| \sqrt{f(1-f)}$\cite{herzog2022} and then Eq.~\eqref{eq.ds_qm} is given by
\begin{equation}
    \sqrt{\mathrm{det} D_s} = \frac{4 N_{of} \sqrt{\mathrm{det} M^{\mathrm{min}}}}{(2 \pi)^{D-1} N_f |U|} \Delta^2. \label{eq.ds_fb_upc}
\end{equation}
Intriguingly, even when this relation only holds for isolated flat bands with UPC at zero temperature, our numerical results give evidence that superfluid weight can be proportional to $\Delta^2$ also in some nonisolated flat bands at non-zero temperatures. 

We have also determined the order parameters of the lattices without UPC, i.e. the x-directional and decorated extensions (see Supplementary Note 5). In the x-directional extension $|\Delta_B|$ and $|\Delta_C|$ behave approximately as $\Delta(T) = \Delta(0)\sqrt{1-(T/T_c)^b}$ for $b$ between 4 and 5, while $\Delta_C$ behaves as the order parameters of the lattices with UPC. In the decorated extension, $\Delta_C$ and $\Delta_B$ behave as the order parameters of the lattices with UPC, while $\Delta(T) = \Delta(0)\sqrt{1-(T/T_c)^3}$ agrees with the behavior of $\Delta_D$. For these lattices without UPC the temperature behavior of the superfluid weight is not straightforwardly proportional to $\Delta^2$.

\subsection{Mean-field comparison}

We have tested how important the fluctuations included in the DMFT method are by comparison to mean-field calculations (see Supplementary Note 4). We find that also the mean-field results can convey our message of flat-band ratio and zero temperature superfluid weight, or Eq.~\eqref{eq.ds_qm}, being able to predict the BKT temperature. In most cases, this works even better within mean-field theory (for example in the case of the twice-extended Lieb lattice) since there the critical temperatures can be predicted with  $T_C \approx |U| N_f/(4 N_{of})$ for all lattices; however, we emphasize that DMFT shows this is not the case when fluctuations are included.  

The mean-field and DMFT results are in good quantitative agreement in the low-interaction, low temperature limit in many of the models we consider, which is expected since local thermal and quantum fluctuations taken into account in DMFT but not in mean-field theory typically have a small impact in this regime. At higher temperatures, the thermal fluctuations included in DMFT decrease critical temperatures compared to mean-field theory. The impact of thermal fluctuations can vary from one model to the other, so even qualitative predictions from mean-field theory can be slightly inaccurate (for instance, in the "Two-dimensional extensions" subsection of the Results section, the twice-extended lattice has a lower $T_c$ than the Lieb and connected lattices, which is not predicted by mean-field theory). Similarly, when the interaction is increased, DMFT estimates lower critical temperatures and smaller pairing gaps than mean-field theory. For the fairly low interactions we consider here, it seems that mean-field predictions often hold at least on a qualitative level (see Supplementary Note 4).

In some lattices, DMFT and mean-field theory disagree even in the low-temperature, low-interaction limit where they would typically agree. This is notably the case in systems where some orbitals can be disconnected from the lattice while keeping the original Lieb lattice intact. In the three-dimensional extended Lieb, in the extreme limit where the sublattice $D$ is completely disconnected from the others, mean-field theory predicts a nonzero and fairly large order parameter $\Delta_D$, whereas it is zero in DMFT because fluctuations are properly taken into account. When $d$ is increased and $D$ is connected to the rest of the lattice, qualitative differences between mean-field and DMFT subsist. This has the following 
consequence: in mean-field theory, the critical temperature has an optimum in the quasi-two-dimensional regime, i.e. for small $d$, see Supplementary Note 4 for details. However, this result is not present at the DMFT level. Thus, beyond-mean-field calculations are needed to produce reliable results for the different extensions. Furthermore, DMFT calculations confirm that the flat-band ratio and the zero temperature superfluid weight can be used to predict the $T_c$ and $T_{BKT}$ even when some of the fluctuations on the systems are accounted for.

\section{Conclusions}

We investigated the critical temperature for superconductivity in several two- and three-dimensional extensions of the Lieb lattice with DMFT. While the superconducting properties for these extensions differ in various ways, we found that the flat-band ratio, i.e. the ratio of the number of flat bands at the Fermi energy to the total number of bands (or orbitals), plays a role in determining the critical temperature. For all of the models considered, except the 2D and 3D twice-extended ones, the temperature at which the superconducting order parameters and the superfluid weights obtained from the DMFT calculations vanish increases when the flat-band ratio is increased. In a sense, this result can be expected, since the number of states at the Fermi energy is increased with the number of flat bands. 

The importance of quantum geometry in the zero-temperature superfluid weight of flat bands is well-established and has been verified, for instance, by quantum Monte Carlo and DMFT computations~\cite{Julku2016,Peri2021,Herzog-Arbeitman2022,Hofmann2023}. In this $T=0$ limit, it is known that both the flat-band ratio and the minimal quantum metric play a role. Our results show that in this limit with small interaction strengths, the simple mean-field analytical result for the superfluid weight agrees with DMFT calculations even when the isolated flat band and uniform pairing conditions used in its derivation are poorly met. Furthermore, we find that with larger interaction strengths the analytical mean-field result agrees qualitatively with the DMFT calculations.  A higher flat-band ratio can, in principle, result in a higher zero-temperature superfluid weight, but the quantum metric becomes zero when the ratio is brought to its maximal value of one (meaning that the system consists of only degenerate flat bands). 

The zero-temperature superfluid weight typically only gives an upper bound for the critical temperature. The precise determination of $T_{BKT}$ requires knowledge of the superfluid weight also at nonzero temperature. However, we find that the zero-temperature superfluid weight (obtained from either DMFT or mean-field calculations) together with the flat-band ratio can qualitatively predict $T_{BKT}$ across different lattices: increasing the zero-temperature superfluid weight and/or the flat-band ratio increases $T_{BKT}$. In isolated flat-band lattices, the flat-band ratio and the minimal quantum metric can qualitatively predict the DMFT zero-temperature superfluid weight. This provides an even simpler way to qualitatively predict $T_{BKT}$ only from the flat-band ratio and the minimal quantum metric which are non-interacting properties of the system.

Our results point to a potential method to design flat band systems with increased critical temperatures: the addition of orbitals resulting in an increased number of flat bands at the Fermi energy. The possible enhancement of the critical temperature is especially high if this extension adds a third dimension to a two-dimensional system, since reaching the BKT transition is no longer required. In our example case, this would be at the expense of the critical current at low temperatures since there the two-dimensional superfluid weight is larger. 

Flat band models have been realized in ultracold atom experiments~\cite{Jo2012,Li2024,Li2022,Leykam2018}, including the Lieb lattice~\cite{Taie2015,Taie2020}. The high tunability of these systems makes them good candidates for the realization of lattice extensions. Moir\'e and other two-dimensional quantum materials~\cite{Andrei2020TBG,Andrei2021,Wang2023,Torma2022,Efetov2024,Tian2023} offer unique possibilities for tuning the number of flat bands, orbitals in the (moir\'e) unit cell, and the quantum metric. One more interesting area of further study is stoichiometric three-dimensional flat band materials~\cite{Regnault2022} where the number of relevant orbitals, flat bands and their quantum metrics can be obtained from ab-initio calculations and could help searches of materials with high critical temperatures and critical currents.

\section{Methods}
\label{sec:Methods}

The usual mean-field treatment, where the interaction is replaced by an interaction with a mean-field, does not account for any of the fluctuations in the system. To include fluctuations at the local level, we use dynamical mean-field theory (DMFT), where the full lattice problem is mapped to an Anderson impurity model, in which the impurity is coupled to a bath of non-interacting particles~\cite{hettler2005cluster, marcelo1996dmft, hettler200dca, Haule2006}. We use real-space DMFT where each site in the unit cell is mapped to its own impurity problem. This allows us to account for fluctuations inside these orbitals that the usual mean-field approximation neglects. The essential approximation of DMFT is that the self-energy of the full lattice model is approximated by the self-energies obtained by solving the impurity problems. This results in a self-energy that is local to each unit cell and each orbital and varies only within unit cells but not between them, i.e., $\bm{\Sigma}_{i, \alpha j, \beta}(i\omega_n)\approx \delta_{i j}\delta_{\alpha \beta}\bm{\Sigma}_{\alpha}(i\omega_n)$, where $i, j$ and $\alpha, \beta$ are the indices of the unit cells and orbitals respectively and $\omega_n = \pi(2n+1)T$ are the fermionic Matsubara frequencies and $T$ is the temperature.

Here we briefly describe how the DMFT works in the multi-band case described above. The full Green's function can be obtained from the non-interacting Green's function $\bm{G}^0_{i}(i\omega_n)$ and the self-energy with the Dyson equation 
\begin{equation}
    \bm{G}_{i}(i\omega_n) = \frac{1}{\bm{G}^0_{i}(i\omega_n)^{-1} -\bm{\Sigma}_i(i\omega_n)}.
\end{equation}
Since the self-energy is local to the unit cell, the non-local full Green's function is just the non-interacting Green's function, i.e., $\bm{G}_{i\neq0}(i\omega_n) = \bm{G}^0_{i\neq0}(i\omega_n)$. After a Fourier transformation, the Dyson equation for the local Green's function can be rewritten in momentum space as
\begin{equation}
    \bm{G}_{loc}(i\omega_n) = \frac{1}{N_k}\sum_{\bm{k}}\frac{1}{\bm{G}^0_{\bm{k}}(i\omega_n)^{-1} - \bm{\Sigma}(i\omega_n)},
    \label{eq:G local from K}
\end{equation}
where $N_k$ is the number of unit cells, i.e momentum points and $\bm{G}^0_{\bm{k}}(i\omega_n)^{-1}$ is the non-interacting Green's function in momentum space. The self-energy is a constant in momentum space because it is local to the unit cell in the real space. From this local Green's function and the self-energy, we can obtain the dynamical Weiss mean-field, which is essentially the non-interacting local Green's function
\begin{equation}
    \bm{\Gamma}(i\omega_n) = \frac{1}{\bm{G}_{loc}(i\omega_n)^{-1}+\bm{\Sigma}(i\omega_n)}.
    \label{eq:weiss field}
\end{equation}
Then $\bm{\Gamma}(i\omega_n)$ is used to define the Anderson impurity problem which we solve with the interaction expansion of continuous time Monte Carlo solver (CT-INT)~\cite{MonteCarlo2011, MonteCarlo2016}. We solve the self-energies and Green's functions self-consistently and then obtain the order parameters for each orbital from the anomalous components of the local Green's functions according to equation~\eqref{eq:order parameter}.

In two-dimensional systems, the transition to superconductivity occurs at the Berezinskii-Kosterlitz-Thouless (BKT) temperature $T_{BKT}$ which can be obtained from the relation~\cite{Berezinsky1971, Thouless1973, Nelson1977}
\begin{equation}
    T_{BKT} = \frac{\pi}{8}\sqrt{\mathrm{det}[D^s(T_{BKT})]}.
    \label{eq:Tbkt}
\end{equation}
The superfluid weight is therefore a central quantity in this study.
In DMFT, the superfluid weight can be obtained from the current density $\expval{j_\mu}$ induced by a constant vector potential $A_\nu$ with the linear response formula \cite{liang2017}
\begin{equation}
    \expval{j_\mu} =  -D^s_{\mu \nu} A_\nu.
    \label{eq:superfluidweightDMFT}
\end{equation}
Here, the constant vector potential $\bm{A}$ enters the hopping staggering parameters by the usual Peierls substitution
\begin{align}
    t'_{i \alpha j \beta}(\bm{A}) &= e^{-i \int_{\bm{r}_j+\bm{\delta}_\beta}^{\bm{r}_i+\bm{\delta}_\alpha} \bm{A}(\bm{r}) d\bm{r}} t_{i \alpha j \beta}\nonumber\\
    &= e^{-i (\bm{r}_i+\bm{\delta}_\alpha-\bm{r}_j-\bm{\delta}_\beta)\cdot \bm{A}} t_{i\alpha j \beta},
    \label{eq:HoppingsWA}
\end{align}
where the $\bm{r}_i$ and $\bm{r}_j$ denote the positions of unit cells and $\bm{\delta}_\alpha$ and $\bm{\delta}_\beta$ are the intra unit cell positions of the orbitals $\alpha$ and $\beta$. The current density operator at direction $\mu \in \{x,y,z\}$ can be obtained as the first derivative of the Hamiltonian with respect to a component of vector potential $\bm{A}$ \cite{vanhala2018, huhtinen2022}
\begin{equation}
    j_\mu = \frac{\partial H}{\partial \bm{A}_\mu} = \frac{\partial H^{kin}}{\partial \bm{A}_\mu} = \sum_{i,j} \frac{\partial t_{i,j}(\bm{A})}{\partial \bm{A}_\mu} c_i^\dagger c_j.
\end{equation}
Then, the expectation value of the current operator can be calculated in the momentum space with~\cite{liang2017}
\begin{equation}
    \expval{j_\mu} = \frac{1}{N}\sum_{\bm{k}}M_{\bm{k}}-\frac{1}{N\beta}\sum_{\bm{k}}\sum_{n}M_{\bm{k}}G_{\bm{k}}(i\omega_n),
    \label{eq:kspaceJ}
\end{equation}
where $M_{\bm{k}}$ is the matrix representation of the current operator in $\bm{k}$-space, $G_{\bm{k}}(i\omega_n)$ is the Green's function in momentum and frequency space and $N_k$ is the number of $\bm{k}$ points. 

 Now, the superfluid weight can be obtained using the following procedure: First, a constant small vector potential $\bm{A}=A_\nu$ in direction $\nu \in \{x,y,z\}$ is introduced to the hoppings with the Peierls substitution. Then the Green's functions of the system are calculated with the DMFT algorithm and the induced current in direction $\mu \in \{x,y,z\}$ is obtained with Eq.\eqref{eq:kspaceJ}. Finally, the $\mu, \nu$ component of the superfluid weight can be obtained from Eq.~\eqref{eq:superfluidweightDMFT}. 
 
This procedure of calculating the superfluid weight does not work for every method and system, since a constant vector potential $\vec{A}$ is usually gauge-equivalent to a zero vector potential, which does not produce any supercurrent \cite{liang2017}. However, at the DMFT level, we impose that the anomalous component of the self-energy, including the order parameters, is uniform in space, which breaks this gauge symmetry. Then the constant vector potential produces a phase twist on the order parameters and the off-diagonal anomalous components, which can not be gauged away.

Dynamical mean-field theory only becomes exact in the limit of infinite coordination number, which remains fairly low in the systems we consider. Moreover, our method does not account for potential effects of for instance the vortex-core energy, which can make the vortex unbinding temperature differ from the BKT temperature we determine~\cite{Benfatto2007}. While it likely is better than in mean-field theory, the accuracy of the critical temperatures DMFT predicts might still be limited by these approximations. 

\section{Data availability}
The data generated and analyzed in this study are available from the
corresponding author on a reasonable request.

\section{Code availability}
The code used in this study is available upon reasonable request from the corresponding author.

%Used to add a title for the references section
\def\bibsection{\section*{\refname}} 
\bibliography{main_refs}

\section{Acknowledgments}
The authors wish to acknowledge CSC – IT Center for Science, Finland, for computational resources. R.P.~acknowledges financial support from the Fortum and Neste Foundation. K.-E.H.~is supported by an ETH Zurich Postdoctoral Fellowship. This work was supported by the Research Council of Finland (former Academy of Finland) under project number 339313 and by Jane and Aatos Erkko Foundation, Keele Foundation and Magnus Ehrnrooth Foundation as part of the SuperC collaboration.

\section{Author contributions}
PT initiated and supervised the project. RPS did the DMFT calculations. All authors discussed the results. RPS wrote the manuscript together with KEH and PT. 

\section{Competing interests}
The authors declare no competing interests.

\clearpage
\onecolumngrid

\section*{Supplementary Note 1: Band structures }
\label{app:band_gaps}
\begin{figure}[h]
	\centering
	\includegraphics[width=0.7\linewidth]{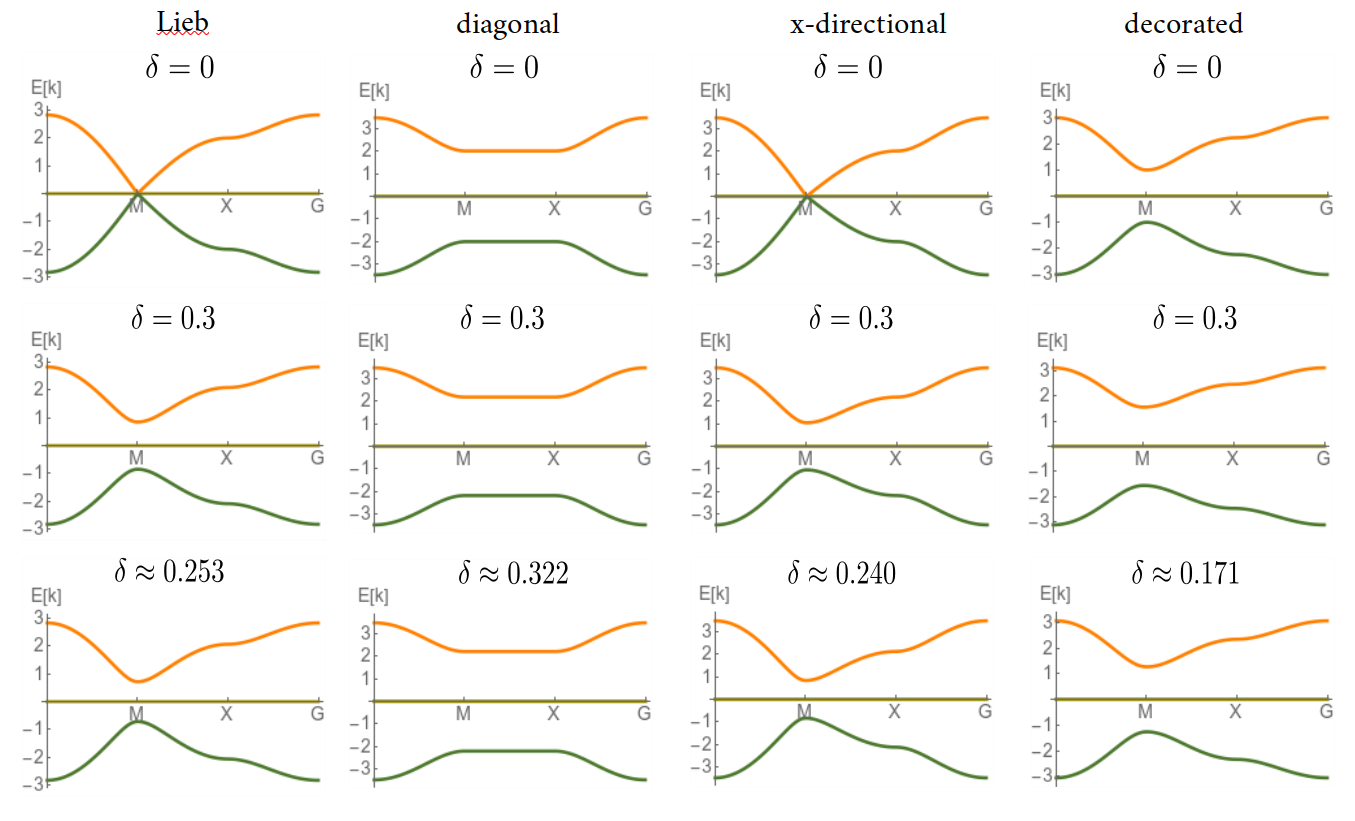}
	\caption{Band structures plotted along the high symmetry lines of the first Brillouin zone for the different versions of the extended Lieb lattice and the original Lieb lattice with hopping staggerings $\delta=0$ (first row), $\delta=0.3$ (second row) and with such $\delta$ that $\mathrm{det} M^{\rm min}=1.3$ (third row). The Lieb lattice has one flat band at zero energy for all $\delta$, and the different versions all have two degenerate flat bands at zero energy.}
	\label{fig:2D_bands_extended}
\end{figure}

The first and second rows of Fig.~\ref{fig:2D_bands_extended} show the band structures of the different versions and the original Lieb lattice with hopping staggering parameters $\delta = 0$ and $0.3$, respectively. The original Lieb lattice and the x-directional version have band touchings at the high-symmetry point $M$ when $\delta = 0$, which are opened with $\delta = 0.3$. The band structures of the diagonal and decorated versions have no band touchings even with $\delta=0$. Furthermore, both dispersive bands of the diagonal Lieb lattice have a flat band portion along the high symmetry line from point $M$ to point $X$. The last row of Fig.~\ref{fig:2D_bands_extended} shows that the diagonal and the decorated versions have the largest band gaps when the lattices have the same determinant of the minimal quantum metric of $\mathrm{det} M^{\rm min}=1.3$. We also note that the band structures of the original Lieb and the x-directional version look similar in this case. Figure~\ref{fig:quantummetric} shows the quantum metric distributed over the Brillouin zone in the different lattices. This helps us understand why certain lattices have higher superfluid weight than others. We note that the quantum metric can have a large effect when integrated over the Brillouin zone, even when there is a relatively large (compared to $U$) band gap to the next band(s).

\section*{Supplementary Note 2: Quantum metric prediction}
\label{app:QMpred}

\begin{figure}[h]
	\centering
	\includegraphics[width=1\linewidth]{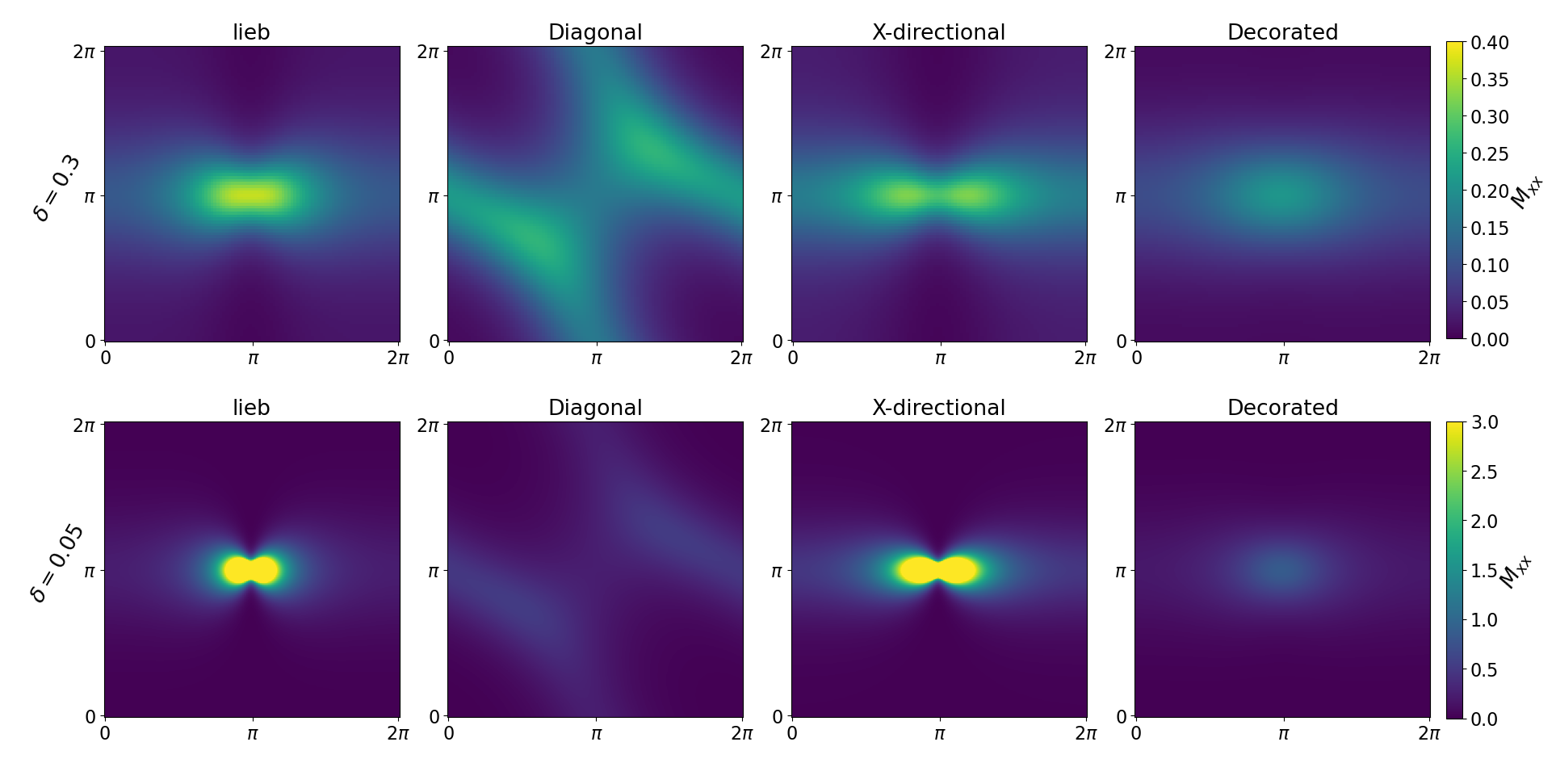}
	\caption{The minimal quantum metric ($M_{xx}$) of the flat band(s) in the first Brillouin zone for the different versions of the extended Lieb lattice with hopping staggerings $\delta = 0.3$ (first row) and with $\delta = 0.05$ (second row). }
	\label{fig:quantummetric}
\end{figure}

\begin{figure}[h]
	\centering
	\includegraphics[width=1\linewidth]{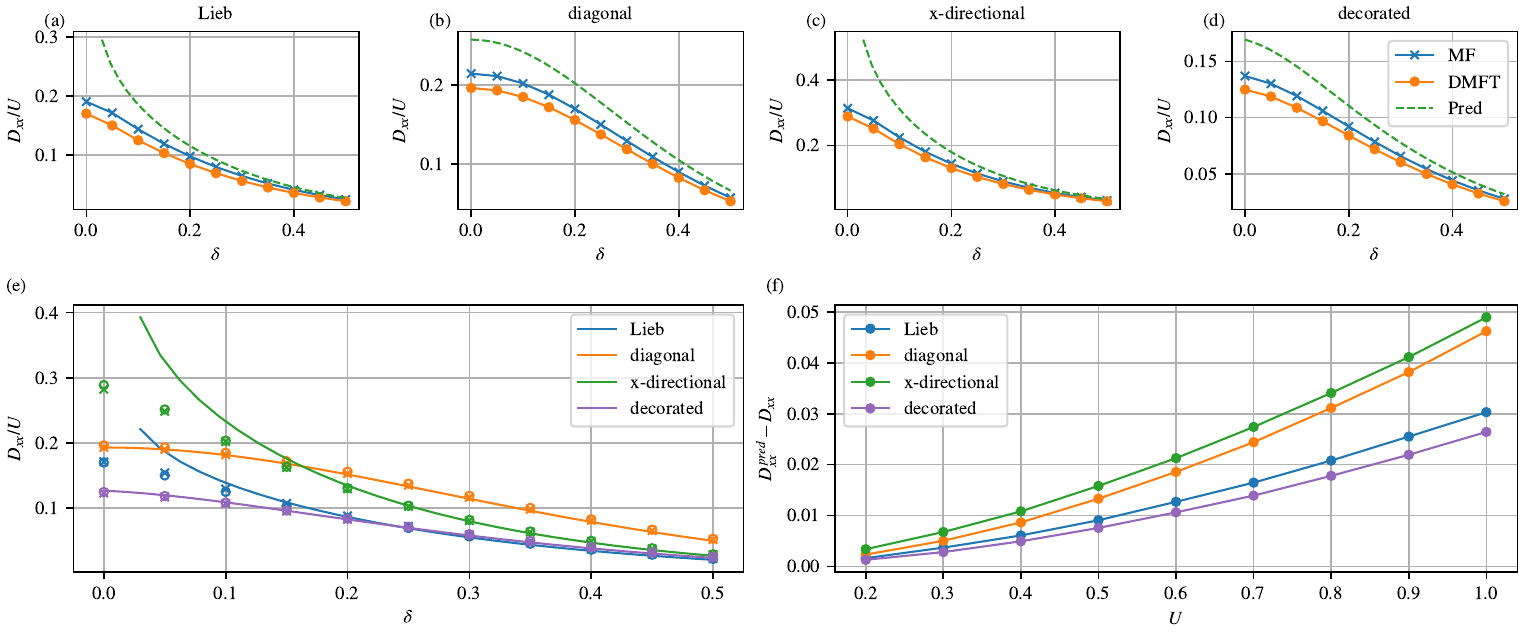}
	\caption{ a-d) The $D_{xx}$ component of the superfluid weight at temperature $T=0.005$ with interaction strength $|U | = 1$ for the original Lieb lattice and the different versions of the two-dimensional extended Lieb lattices obtained from DMFT, mean-field calculations, and Eq. (7).  e) The $D_{xx}$ component of the superfluid weight from DMFT and mean-field calculations at $T=0.005$ with $|U|=1$, and from the minimal quantum metric prediction Eq. (7), for the original Lieb lattice and the different versions of the two-dimensional extended Lieb lattices. The points give the DMFT superfluid weights, the lines give the predictions calculated scaled by a factor of $\frac{3}{4}$ and the crosses give the mean-field superfluid weights scaled by a factor of $\frac{9}{10}$. f) Difference between the superfluid weight predicted from the minimal quantum metric at zero temperature and the DMFT superfluid weight as a function of the interaction strength $|U|$ with a hopping staggering $\delta = 0.2$. }
	\label{fig:Ubehaviour}
\end{figure}

From Figs.~\ref{fig:Ubehaviour} a-d)  we find that both the mean-field superfluid weight and the superfluid weight predicted with the equation (7) in the main text overestimate the DMFT superfluid weight for all versions of the extended Lieb lattice and the original Lieb. The overestimation is larger for superfluid weights predicted by Eq. (7) compared to the numerical mean-field calculation including the Hartree terms. Despite these quantitative differences, both the mean-field and Eq. (7) superfluid weights seem to have the same qualitative behavior as the DMFT results. Fig.~\ref{fig:Ubehaviour} e) confirms this: when the prediction of Eq. (7) is multiplied by a factor of $\frac{3}{4}$ and the mean-field results by a factor of $\frac{9}{10}$, they match with each other and the DMFT results well for all of the lattices when the band gap is sufficiently large. In other words, with $|U|=1$, Eq. (7) systematically overestimated the DMFT superfluid weight by a factor of $4/3$, when the band gaps are sufficiently large, and the mean-field overestimates the DMFT results by a factor of $\frac{10}{9}$. Figure~\ref{fig:quantummetric} f) shows that the difference between $D_{xx}$ predicted from the minimal quantum metric and DMFT superfluid weight at $T=0.005$ increases as a function of $|U|$. These results indicate that when the interaction strength is increased both the Eq. (7) and the mean-field calculations overestimate the superfluid weight compared to the DMFT results but succeeds to predict the qualitative behavior of $D_{xx}$ as a function of $\delta$.

\clearpage

\section*{Supplementary Note 3: Mean-field critical temperature} \label{appendix:MFcritical}

In this section, we compute the temperature $T_{BCS}$ at which the order parameters vanish within mean-field theory in isolated flat bands with uniform pairing. We use a mean-field approximation of the interacting part of the Hubbard model (Eq. (1) of the main text),
\begin{equation}
U\sum_{i\alpha}c_{i\alpha\uparrow}^{\dag} c_{i\alpha\uparrow}^{\vphantom{\dag}} c_{i\alpha\downarrow}^{\dag} c_{i\alpha\downarrow}^{\vphantom{\dag}} \approx \sum_{i\alpha} \left[ \Delta_{\alpha} c_{i\alpha\uparrow}^{\dag}c_{i\alpha\downarrow}^{\dag} + {\rm H.c.} - |\Delta_{i\alpha}|^2/U \right],
\end{equation}
where $\Delta_{i\alpha} = U\langle c_{i\alpha\downarrow} c_{i\alpha\uparrow} \rangle$ is the mean-field order parameter. Assuming that $\Delta_{i\alpha} = \Delta_{\alpha}$ is independent of the unit cell, the Hamiltonian in momentum space can be written in terms of the Bogoliubov-de-Gennes Hamiltonian $H_{\rm BdG}$ as $H=\sum_{\vec{k}}\vec{c}^{\dag}_{\vec{k}} H_{\rm BdG} \vec{c}^{\vphantom{\dag}}_{\vec{k}}$, with
\begin{align}
H_{\rm BdG} &= \begin{pmatrix}
H_{\vec{k}} -\mu & \vec{\Delta} \\
\vec{\Delta}^{\dag} & -H_{-\vec{k}}^*+\mu
\end{pmatrix}, \\
\vec{c} &= (c_{\vec{k}1\uparrow}^{\vphantom{\dag}},\ldots,c_{\vec{k}N_o\uparrow}^{\vphantom{\dag}},c_{-\vec{k}1\downarrow}^{\dag},\ldots,c_{-\vec{k}N_o\downarrow})^{T}.
\end{align}
Here, $[H_{\vec{k}}]_{\alpha\beta} = \sum_{i} t_{0\alpha,i\beta}
e^{i\vec{k}\cdot(r_{0\alpha}-r_{i\beta})}$, where $r_{i\beta}$ is the
position of the lattice site $i\beta$, and
$[\vec{\Delta}]_{\alpha\beta} =
\Delta_{\alpha}\delta_{\alpha\beta}$. We assume now that the pairing
is uniform throughout the lattice, i.e. $\Delta_{\alpha}=\Delta$, so
that $\vec{\Delta} = \Delta 1$. We also assume time-reversal
symmetry, implying $H_{\vec{k}} = H_{-\vec{k}}^*$, and take $\Delta$
real.

The non-interacting kinetic Hamiltonian $H_{\vec{k}}$ can be
diagonalized as $H_{\vec{k}} 
= \mathcal{G}_{\vec{k}} \epsilon_{\vec{k}}
\mathcal{G}_{\vec{k}}^{\dag}$, where $[\epsilon_{\vec{k}}]_{m,n} =
\epsilon_n(\vec{k}) \delta_{mn}$ contains the band dispersions and
$[\mathcal{G}_{\vec{k}}]_{\alpha n} = \langle \alpha | n_{\vec{k}}
\rangle$ the Bloch functions. We can thus rewrite $H_{\rm BdG}$ as 
\begin{equation}
  H_{\rm BdG} = \begin{pmatrix}
    \mathcal{G}_{\vec{k}} & 0 \\
    0 & \mathcal{G}_{\vec{k}}
  \end{pmatrix}
  \begin{pmatrix}
    \epsilon_{\vec{k}}-\mu & \Delta \\
    \Delta & -\epsilon_{\vec{k}}+\mu
  \end{pmatrix}
  \begin{pmatrix}
    \mathcal{G}_{\vec{k}}^{\dag} & 0 \\
    0 & \mathcal{G}_{\vec{k}}^{\dag}
  \end{pmatrix}.
\end{equation}
The matrix $H_{\rm BdG}$ is now easily diagonalized, and has
eigenvalues and vectors
\begin{align}
  E_n^{\pm}(\vec{k}) &\equiv \pm E_n = \pm\sqrt{(\epsilon_m(\vec{k})-\mu)^2+\Delta^2}, \\
  | \psi_n^+(\vec{k})\rangle &= (u_{n}(\vec{k}) |+\rangle +
  v_n(\vec{k}) |-\rangle )\otimes
  |n_{\vec{k}}\rangle, \\
  | \psi_n^-(\vec{k})\rangle &= (-v_{n}(\vec{k}) |+\rangle +
  u_n(\vec{k}) |-\rangle )\otimes
  |n_{\vec{k}}\rangle, \\
\end{align}
where
\begin{equation}
  u_m(\vec{k}) = \frac{1}{\sqrt{2}} \sqrt{1+\frac{\epsilon_m(\vec{k})-\mu}{E_m(\vec{k})}}, \:\:\:\:\:
  v_m(\vec{k}) = \frac{1}{\sqrt{2}} \sqrt{1-\frac{\epsilon_m(\vec{k})-\mu}{E_m(\vec{k})}}.
\end{equation}
The vectors $|\pm\rangle$ are the eigenvectors of the Pauli matrix
$\sigma_z$ corresponding to eigenvalues $\pm 1$. 
The full BdG Hamiltonian can thus be rewritten in a diagonal form $H =
\sum_{\vec{k}}\vec{\gamma}^{\dag}_{\vec{k}} \vec{E}_{\vec{k}} \vec{\gamma}^{\vphantom{\dag}}_{\vec{k}} + C$, where $C$ is a scalar
which does not impact the following calculations and
\begin{align}
  \vec{E}_{\vec{k}} &= {\rm diag}(E_{1}^+(\vec{k}),\ldots E_{N_o}^+(\vec{k}), E_{1}^-(\vec{k}),\ldots,
  E_{N_o}^-(\vec{k})), \\
  \vec{\gamma}_{\vec{k}} &=
  (\gamma_{\vec{k}1+},\ldots,\gamma_{\vec{k}N_o+},\gamma_{\vec{k}1-},\ldots,\gamma_{\vec{k}N_o-})^T,\\
  c_{\vec{k}\alpha\uparrow} &= \sum_{n} \langle \alpha |
    n_{\vec{k}}\rangle\left[ u_n(\vec{k})  \gamma_{\vec{k}n+} - v_n(\vec{k}) \gamma_{\vec{k}n-} \right], \\
  c_{-\vec{k}\alpha\downarrow}^{\dag} &= \sum_{n} \langle \alpha |
    n_{\vec{k}}\rangle\left[ v_n(\vec{k})
     \gamma_{\vec{k}n+} + u_n(\vec{k})  \gamma_{\vec{k}n-}\right]. 
\end{align}

The order parameters can be solved from the gap equation
\begin{align}
  \Delta_{\alpha} &= -\frac{|U|}{N_{k}}\sum_{\vec{k}} \langle
  c_{-\vec{k}\alpha\downarrow} c_{\vec{k}\alpha\uparrow} \rangle \\
  &= -\frac{|U|}{N_{k}}\sum_{\vec{k}} \sum_n |\langle \alpha |
  n_{\vec{k}}\rangle|^2 \left[ u_n(\vec{k})v_n(\vec{k}) \langle \gamma_{\vec{k}n+}^{\dag}
    \gamma_{\vec{k}n+}^{\vphantom{\dag}}\rangle - u_n(\vec{k})v_n(\vec{k}) \langle
    \gamma_{\vec{k}n-}^{\dag} \gamma_{\vec{k}n-}^{\vphantom{\dag}} \rangle \right]
  \\
  &= \frac{|U|\Delta}{2N_{k}} \sum_{\vec{k}}\sum_{n} |\langle \alpha |
  n_{\vec{k}} \rangle|^2 \frac{\tanh(\beta E_n(\vec{k})/2)}{E_n(\vec{k})}. 
\end{align}
Because we assume uniform pairing,
\begin{align}
  &\Delta = \frac{1}{N_o}\sum_{\alpha} \Delta_{\alpha} =
  \frac{|U|\Delta}{2N_{k}N_o} \sum_{\vec{k}}\sum_n\frac{\tanh(\beta
    E_n(\vec{k})/2)}{E_n(\vec{k})}.\label{eq.gap}
\end{align}

We now consider a set $\mathcal{S}$ of $N_f$ degenerate flat bands at energy
$\epsilon_{\overline{n}}$. Then Eq.~\eqref{eq.gap} becomes
\begin{equation}
  1 =
  \frac{|U|N_f}{2N_o}\frac{\tanh(\beta\sqrt{(\epsilon_{\overline{n}}-\mu)^2+\Delta^2}/2)}{\sqrt{(\epsilon_{\overline{n}}-\mu)^2
  + \Delta^2}} +
  \frac{|U|}{2N_{k}N_o}\sum_{\vec{k}}\sum_{n\notin\mathcal{S}}
  \frac{\tanh(\beta E_n(\vec{k})/2)}{E_n(\vec{k})}.  
\end{equation}
Setting $\mu=\epsilon_{\overline{n}}$ and taking the limit $T\to
T_{BCS}$ where $|\Delta|\to 0^+$ yields the critical temperature for a half-filled flat band
\begin{equation}
T_{\rm BCS} = \frac{|U|N_f}{4N_o}  +
  \frac{|U|}{2N_{k}N_o} \sum_{\vec{k}}\sum_{n\notin\mathcal{S}}
  \frac{\tanh(|\epsilon_n(\vec{k})-\epsilon_{\overline{n}}|/(2T_{\rm
      BCS}))}{|\epsilon_n(\vec{k})-\epsilon_{\overline{n}}|/T_{\rm BCS}}.  \label{Tcestimate}
\end{equation}
If the flat band is isolated so that
$|\epsilon_n(\vec{k})-\epsilon_{\overline{n}}|$ is very large at all
momenta, the second term can be ignored and we obtain a critical
temperature proportional to the flat-band ratio,
\begin{equation}
  T_{\rm BCS} \approx \frac{|U|N_f}{4N_o}.
\end{equation}

This result can be generalized to models where the pairing is uniform only at those orbitals where the flat band states reside, and vanishingly small at other sites. In such cases, $N_o$ should be replaced by the number of orbitals at which the flat band states reside, $N_{of}$. In the two-dimensional Lieb lattice, where the flat band states reside at the $B/C$ sites, this gives an estimate of $|U|/8$ for $T_{BCS}$ in the isolated band limit. In the three-dimensional once-extended extension, where the states of the two flat bands reside at the $B/C/D$ sites, the estimate for an isolated band is $|U|/6$. As can be seen from Fig.~\ref{fig:Lieb_extended_Mf}, this can be close to the mean-field $T_{BCS}$ even when the bands are not isolated. The critical temperature obtained from DMFT is lower than the mean-field estimate.

One can make further estimates by assuming that the bands other than the flat bands of interest are (nearly) flat and separated from them by a small gap, i.e., $|\epsilon_n(\vec{k})-\epsilon_{\overline{n}}| \simeq \delta$ where $\delta$ is a constant small compared to $|U|$. This allows to Taylor expand Equation~\eqref{Tcestimate} to second order in $\delta$. This gives
\begin{eqnarray}
T_{BCS} - \frac{|U|}{4} \simeq -\frac{|U|}{T_{BCS}^2} \frac{\delta^2 N_{other}}{48 N_o}, \label{Tcapproximate}
\end{eqnarray}
where $N_{other}$ is the number of the bands other than the flat bands of interest. We see that for small $\delta$, $T_{BCS}$ will be
\begin{equation}
T_{BCS} \sim \frac{|U|}{4} ,
\end{equation}
which can be understood as the optimal case for the uniform pairing condition, since the flatness of the other bands maximizes the $\tanh$-containing term in Equation (\ref{Tcestimate}). This is only the temperature at which the mean-field order parameters vanish, and does not take into account whether the superfluid weight is nonzero. In fact, in a system with only degenerate flat bands, the quantum metric of the degenerate band vanishes and so does the superfluid weight. However, once some of the bands are separated from the flat bands of interest by a gap $\delta$, a nonzero quantum metric and therefore superfluid weight become possible. 

Using $T_{BCS} \sim \frac{|U|}{4}$ on the right hand side of Equation (\ref{Tcapproximate}) gives an estimate how much a finite small value of $\delta$ decreases $T_{BCS}$ from the optimum $|U|/4$:
\begin{equation}
 T_{BCS} \sim  \frac{|U|}{4} - \frac{N_{other}}{3 N_o}\frac{\delta^2}{|U|} .
\end{equation}

\clearpage

\section*{Supplementary Note 4: Mean-field results}
\label{app:MF results}

\begin{figure}[h]
	\centering
	\includegraphics[width=\linewidth]{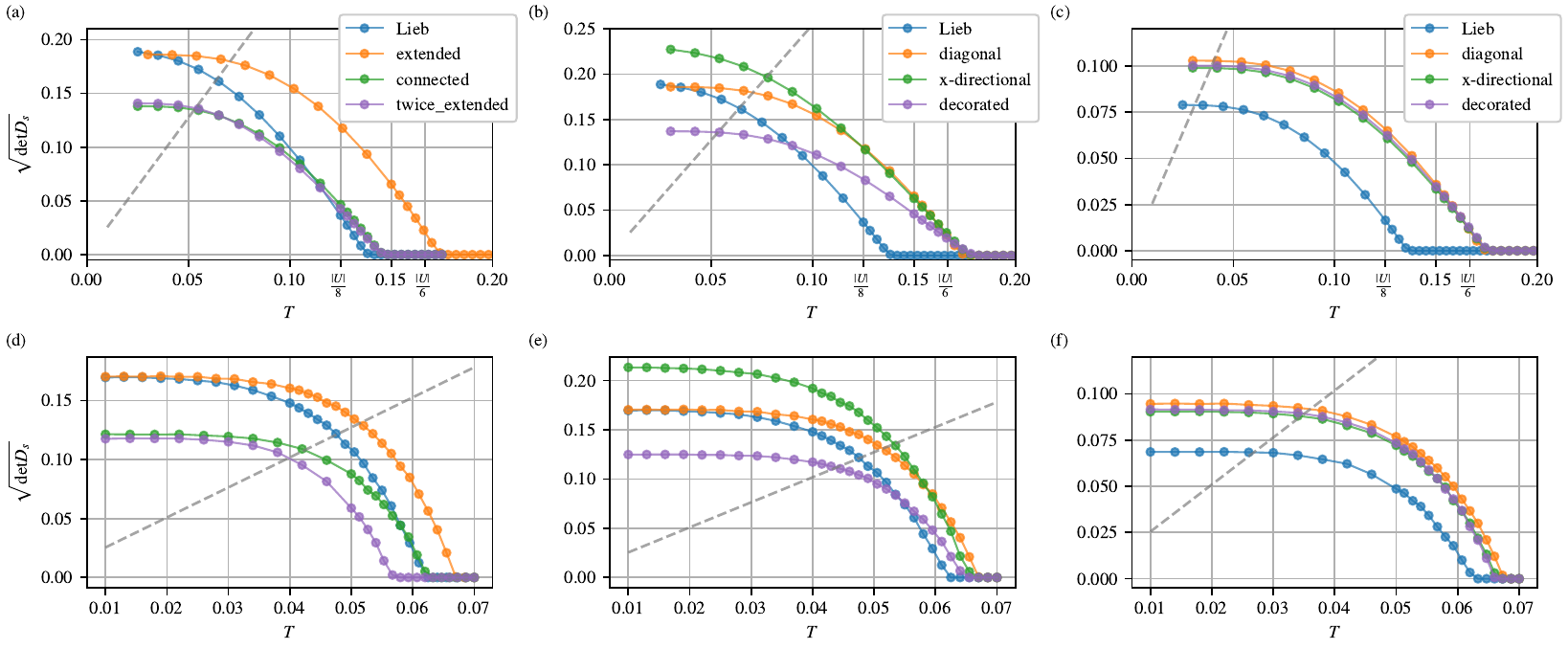}
	\caption{a) and d) give the determinant of the superfluid weight $\sqrt{\mathrm{Det}\, D_s}$ for different extensions of the Lieb lattice shown in Fig. 1 a), calculated with a) the mean-field theory and d) the DMFT. b) and e) give $\sqrt{\mathrm{Det}\, D_s}$ for different versions of the extended Lieb lattice and the original Lieb (see Fig. 1 b)) for hopping staggering $\delta = 0$ calculated with b) mean-field and e) DMFT. c) and f) give $\sqrt{\mathrm{Det}\, D_s}$ for the different versions of the extended Lieb with such $\delta$ that all lattices have the same quantum metric $M=1.3$ calculated with c) mean-field and f) DMFT. In all panels, the interaction strength is $|U|=1$.}
 
	\label{fig:S4}
\end{figure}
The decision to use DMFT calculations in this study was motivated by the quantum fluctuations that it manages to account for. One can also perform similar calculations with mean-field theory and obtain the order parameters and superfluid weights. However, in this case, all of the quantum fluctuations in the system are neglected. In this section, we investigate how mean-field results compare to the DMFT results at non-zero temperatures.

In figure \ref{fig:S4}, we show the superfluid weights calculated with both mean-field (top panels) and DMFT (bottom panels) for the systems considered in Figures 3 b), 4 b) and d) in the main text. The mean-field superfluid weights consistently vanish at temperatures approximately 2.5 times larger than in DMFT. At the DMFT level, the local fluctuations are included, reducing the critical temperatures of the superfluid weight. Fig.~\ref{fig:S4} a) shows that the lattices with a flat-band ratio of $1/2$ (Lieb, connected and twice-extended) have the same mean-field $T_c$, which is close to the isolated flat-band prediction $T_c \approx |U|/8$. Furthermore, the mean-field $T_c$ of the extended Lieb lattice, which has a flat-band ratio of $2/3$, is close to the isolated flat-band prediction $T_c \approx |U|/6$. The flat-band ratio corresponds to the mean-field critical temperatures for all the lattices, including the twice-extended lattice, however, the DMFT results have shown this is not the case when fluctuations are taken into account. 

All the different versions of the extended Lieb lattice (see Fig. 1 b) of the main text) have the same mean-field $T_c$ close to $|U|/8$, shown in Figs.~\ref{fig:S4} b) and c). While DMFT results showed only a marginal increase in $T_c$ for the various extended Lieb lattice versions compared to the original Lieb lattice, the mean-field results have a notably larger difference. Figs.~\ref{fig:S4} b) and c) show that, as in the DMFT results, also at the mean-field level the combination of the zero-temperature superfluid weight and the flat-band ratio can predict the behavior of $T_{BKT}$ across different lattices. Furthermore, when the flat bands are sufficiently isolated $T_{BKT}$ can be predicted from the minimal quantum metric and the flat-band ratio: in Fig.~\ref{fig:S4} c) all the different versions of the extended Lieb lattice have approximately equal $T_{BKT}$ while having the same flat-band ratio and minimal quantum metric. This is expected in mean-field theory, since the superfluid weight in a perfectly isolated flat band with uniform pairing is given by Eq. (9) of the main text, and the gap equation reduces to $|\Delta(T)| = |U|N_{f}/(2N_{o}) {\rm tanh}(|\Delta(T)|/(2T))$, so that $T_{BKT}$ should be equal for lattices with the same flat band ratio and quantum metric.

\begin{figure*}
	\centering
	\includegraphics[width=\linewidth]{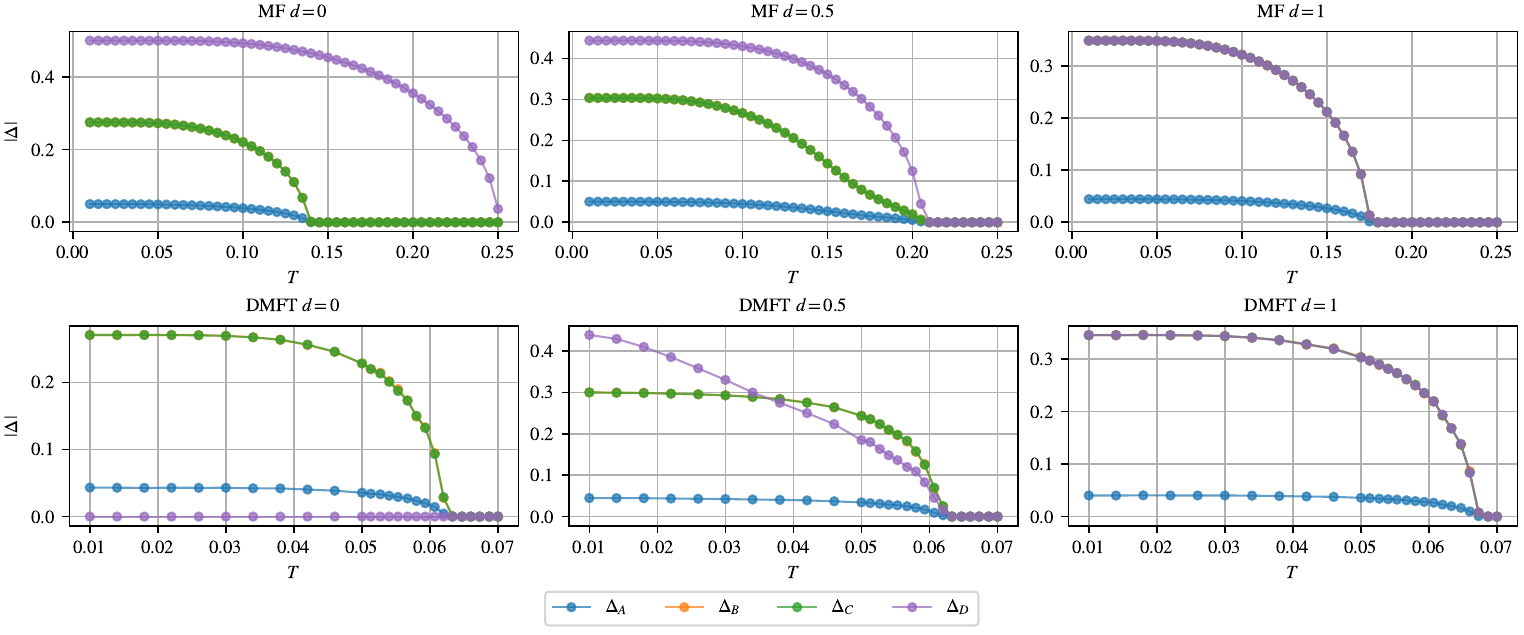}
	\caption{Absolute values of mean-field (top row) and DMFT (bottom row) order parameters as a function of temperature for the 3D extended Lieb lattice with different values of the z directional hopping $d$. The $|\Delta_B|$ and $|\Delta_C|$ are equal in all figures.  }
	\label{fig:Lieb_extended_Mf}
\end{figure*}

Fig.~\ref{fig:Lieb_extended_Mf} shows the absolute values of the order parameters for the three-dimensional extended Lieb lattice calculated with both the mean-field theory and DMFT. We have introduced a hopping parameter $d$ that gives the hopping strengths in z-direction, i.e, to the additional site $D$ (see Fig. 1 b) of the main text). When $d=0$, the system consisted of stacked two-dimensional Lieb lattices with disconnected orbitals between them and when $d=1$, we have the three-dimensional extended Lieb lattice. Using this parameter, we can tune our system continuously from the two-dimensional Lieb to the three-dimensional extended Lieb lattice.

The mean field and the DMFT order parameters shown in Fig.~\ref{fig:Lieb_extended_Mf} are both qualitatively and quantitatively different. First, the critical temperatures of these order parameters are significantly larger than in the DMFT results. More interesting is that now the mean-field critical temperature for the order parameters of the flat-band sites $B$ and $C$, with $d = 0.5$ ($T_c \approx 0.21$) is larger than with $d = 0$ ($T_c \approx 0.14$) or with $d = 1$ ($T_c \approx 0.17$). This behavior is distinct from the DMFT results, where the critical temperature with $d=0.5$ is between the $d=0$ and $d=1$ critical temperatures. Thus, the mean-field behavior is clearly different from that of the DMFT results.

This behavior can be explained by the order parameter of the disconnected D site ($\Delta_D$) in the $d = 0$ case. When $d = 0$, $\Delta_D$ is zero in the DMFT results but nonzero (and large) in the mean-field results. This shows that in the absence of fluctuations, i.e., at the mean-field level, Cooper pairing is possible in the disconnected $D$ orbitals. Furthermore, $\Delta_D$ has a larger critical temperature than other orbitals when $d = 0$. For small but non-zero $d$ these strong Cooper pairs in the $D$ sites with large critical temperatures seem to enhance Cooper pairing in other orbitals, which raises the critical temperatures of these orbitals. This results in the mean-field critical temperature being enhanced with $d = 0.5$ compared to the $d=0$ and $d=1$ cases. In this 3D case, the parameter $d$ can be used to tune the lattice from two to three dimensions, and thus, the mean-field critical temperature is maximized not in the two or three-dimensional cases but rather between them in the quasi-two-dimensional regime. However, this large order parameter at the disconnected $D$ site is only an artifact of the mean-field approximation and not physical since it is not present in the more precise DMFT calculations. We find that similar behavior can also be seen in two-dimensional lattices, where one can disconnect some orbitals while still keeping the original Lieb lattice. Such lattices include, for example, the diagonally extended and decorated lattices.

\clearpage
\section*{Supplementary Note 5: Qualitative behavior of the order parameters}
\label{app:scaled_delta}
\begin{figure*}[ht]
	\centering
	\includegraphics[width=\linewidth]{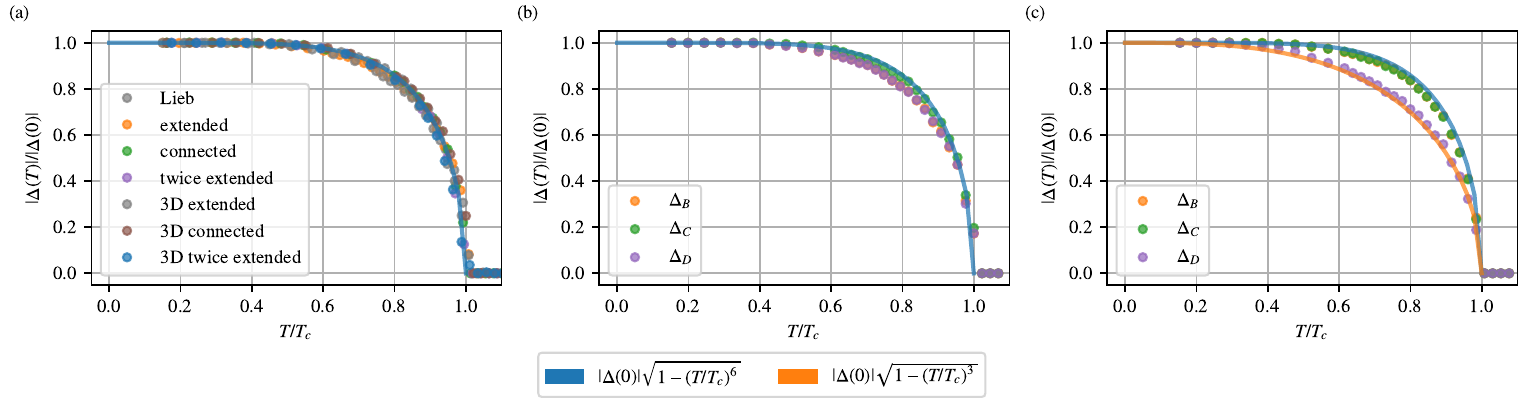}
	\caption{a) The scaled order parameter of the orbitals at which flat-band states reside for all the lattices with UPC with $|U| = 1$. b) (c)) The scaled order parameters of the x-directional (decorated) versions with $|U|=1$. }
	\label{fig:scaled_delta}
\end{figure*}

Fig.~\ref{fig:scaled_delta} a) shows that for all of the studied lattices with UPC, the order parameters of the orbitals at which the flat-band states reside agree well with $|\Delta(T)| = |\Delta(0)|\sqrt{1-(T/T_c)^b}$ when $b =6$. Fig.~\ref{fig:scaled_delta} b) shows the order parameters for the x-directional version which does not have UPC. While $|\Delta_C|$ still agrees well with $b = 6$, the temperature behavior of $|\Delta_B|$ and $|\Delta_D|$ is closer to the one obtained with $b = 5$ . In the decorated extension shown in Fig.~\ref{fig:scaled_delta} c), $|\Delta_B|$ and $|\Delta_C|$ behave according to $b = 6$ while for $|\Delta_D|$ the temperature behavior agrees with $b = 3$. We have checked whether the close-to-zero temperature behavior of the superfluid weights, shown in Fig. 7 of the main text, is captured by the well-known exponential dependence predicted by BCS theory, $D_s(T)/D_0(0) = 1 - \sqrt{2\pi\Delta(0)/T} \exp(-\Delta(0)/T)$ but we found a match only extremely close to $T=0$.

\end{document}